\documentclass[12pt]{oxarticle}
\pdfoutput=1
\usepackage{graphicx, float, array, xspace, amscd, amsmath, amsthm, amssymb}
\usepackage[sort&compress, comma, square, numbers]{natbib}

\renewcommand{\l}{\lambda}

\newcommand{\p}{\pi}

\newcommand{\ch}{\chi}

\newcommand{\cO}{\mathcal{O}}

\newcommand{\IP}{\mathbb{P}}

\newcommand{\IZ}{\mathbb{Z}}

\font\eightrm=cmr8 at 8pt
\font\csc=cmcsc10

\newcommand{\beq}{\begin{equation}}
\newcommand{\eeq}{\end{equation}}
\newcommand{\beqnn}{\begin{equation*}}
\newcommand{\eeqnn}{\end{equation*}}
\newcommand{\bea}{\begin{eqnarray}}
\newcommand{\eea}{\end{eqnarray}}
\newcommand{\bean}{\begin{eqnarray*}}
\newcommand{\eean}{\end{eqnarray*}}

\newcommand{\fref}[1]{Figure~\ref{#1}}

\newcommand{\ord}[1]{\cO\kern-2pt\left(#1\right)}

\newcommand{\ee}{\text{e}}
\newcommand{\ii}{\text{i}}

\newcommand{\place}[3]{\vbox to0pt{\kern-\parskip\kern-7pt
                             \kern-#2truein\hbox{\kern#1truein #3}
                             \vss}\nointerlineskip}

\DeclareFontFamily{U}{wncy}{}
\DeclareFontShape{U}{wncy}{m}{n}{<->wncyr10}{}
\DeclareSymbolFont{mcy}{U}{wncy}{m}{n}
\DeclareMathSymbol{\sha}{\mathord}{mcy}{"58}

\newcommand{\capt}[3]{\parbox{#1}{\renewcommand{\baselinestretch}{1.0}
                                                           \caption{\label{#2}\small\it #3}}}

\newcommand{\cy}{Calabi-Yau\xspace}
\newcommand{\cym}{Calabi-Yau manifold\xspace}
\newcommand{\cys}{Calabi-Yau manifolds\xspace}
\newcommand{\K}{K\"ahler\xspace}
\newcommand{\cicy}[2]{\begin{matrix} #1\end{matrix}\!\left[\begin{matrix}#2 \end{matrix}\right]}
\newcommand{\hodgenos}{(h^{11},\,h^{21})}

\renewcommand{\baselinestretch}{1.1}
\numberwithin{equation}{section}
\proofmodefalse
\begin{document}
\pagestyle{empty}

\ifproofmode\underline{\underline{\Large Working notes. Not for circulation.}}\fi

\begin{center}
\null\vskip0.5in
{\Huge Max Kreuzer's Contributions\\[1ex]
to the Study of\\[1ex]
Calabi-Yau Manifolds\\[1in]}
{\csc Philip Candelas\\[0.3in]}
{\it $^1$Mathematical Institute\hphantom{$^1$}\\
University of Oxford\\
24-29 St.\ Giles'\\
Oxford OX1 3LB, UK\\[6ex]}
\vfill{\bf Abstract}
\end{center}
This is a somewhat personal account of the contributions of Max Kreuzer to the study of \cys and has been prepared as a contribution to the Memorial Volume: Strings, Gauge Fields, and the Geometry Behind --- The Legacy of Maximilian Kreuzer, to~be~published by World Scientific.
\newpage
\setcounter{page}{1}
\pagestyle{plain}
Any account  of Max Kreuzer's career in physics must be  bound up with the history of the study of \cys, to which Max contributed at many levels. There were many currents in this study and work was not done in isolation. Work often advances through a series of challenges, and in reaction to other work. In so far as I have myself been involved in some of these researches it is inevitable that I will have to recall some of these projects that were, at times, inextricably linked with Max's work. For this deficiency of the account let me make this single apology.

The story of \cys goes back, of course, to the researches of Calabi and of Yau. Calabi had seen that the first Chern class presented an obstruction to the existence of complex Ricci-flat manifolds and conjectured, for the case of compact \K manifolds, that this was the only obstruction~\cite{Calabi1954, Calabi1957}. Yau had studied General Relativity and had asked the question of whether there could exist nonsingular and nontrivial, source-free solutions to the Einstein equations. Later he became aware that Calabi had formulated this question as a conjecture for the case of compact \K manifolds. Initially Yau believed that there could be no source-free solution to the Einstein equations, for this case, and set about finding counter-examples to the Calabi conjecture. Each of the supposed counter-examples turned out to suffer from subtle fallacies. Given this, Yau set about proving the conjecture, which he was eventually able to do~\cite{Yau:1977ms}. Yau also saw the importance of vector bundles on \cys and, together with Uhlenbeck, proved the existence and uniqueness of Hermitian Yang-Mills connections for stable vector bundles on \K manifolds~\cite{0615.58045, 0678.58041}.
This extended a previous result of Donaldson for projective algebraic surfaces. This result is known as the Donaldson-Uhlenbeck-Yau theorem.

For physicists, the story begins at a  later stage with the realisation that string theory provides, potentially, both a way of achieving the long-sought goal of unifying the fundamental interactions with a quantum theory of gravitation, and that it admits solutions that look like the world that we see. If this is accepted, then it is surprising that there are not more models of elementary particle physics that are based on string theory compactifications. To what extent viable models exist is a subject that is open to discussion. It seems clear, however, that there would be more but for our lack of~technique.

The nature of heterotic string compactification has evolved: we no longer think in terms of the standard embedding but think of more general heterotic vacua. The elements here are, in the first instance, a \cym together with a stable, holomorphic, vector bundle. More generally still, we may need to relax the condition that the manifolds be \K and contemplate what are known as \cys with torsion or manifolds with heterotic structure. An important part of the story, however, involves \cys and it was to their study, a subject about which we were woefully ignorant, that Max dedicated so much work.

\newpage
It is worth recalling that, at the time physicists became interested in these manifolds, there were perhaps six known explicitly. These were
$$
\IP^4[5]~,~~~\IP^5[3,3]~,~~~\IP^5[4,2]~,~~~\IP^6[3,2,2]~,~~~\IP^7[2,2,2,2]
$$
Together with the $Z$-orbifold. The notation denotes a quintic hypersurface in $\IP^4$ a codimension two submanifold of $\IP^5$, defined by two cubics, and so on.
Soon afterwards Yau produced another example~\cite{Tian:1986ic, Tian:1986ie}, one, moreover, that has Euler number $-6$, so of interest, within the context of the standard embedding, for model building. Yau first proposed the manifold
$$
\cicy{\IP^3 \\ \IP^3}{3 & 0 & 1\\ 0 & 3 & 1}^{14,23}\hskip-15pt\raisebox{-12pt}{,}
$$
here the notation refers to a codimension 3 submanifold of $\IP^3{\times}\IP^3$ cut out by 3 polynomials whose bidegrees, in the variables of the two $\IP^3$'s, are given by the columns of the matrix. The numbers appended to the matrix are the Hodge numbers $\hodgenos$. This manifold has Euler number -18 but one can choose polynomials such that there is a freely acting symmetry $\IZ_3$ and the quotient manifold has Euler number -6. The $c_1=0$ condition amounts to the condition that the row sums of the matrix should be one more than the dimension of the embedding space, as also in the five spaces above.
Generalising Yau's construction led to the class of Complete Intersection Calab-Yau (CICY) manifolds, a class of almost 8,000 manifolds (or rather matrices, since a given manifold might be represented in different ways), which seemed like a very general class at the time. I have mentioned the CICY manifolds, which remain a useful source of manifolds, largely to contrast this class with the much bigger classes that came later. 
\begin{figure}[t]
\begin{center}
\includegraphics[width=6.5in]{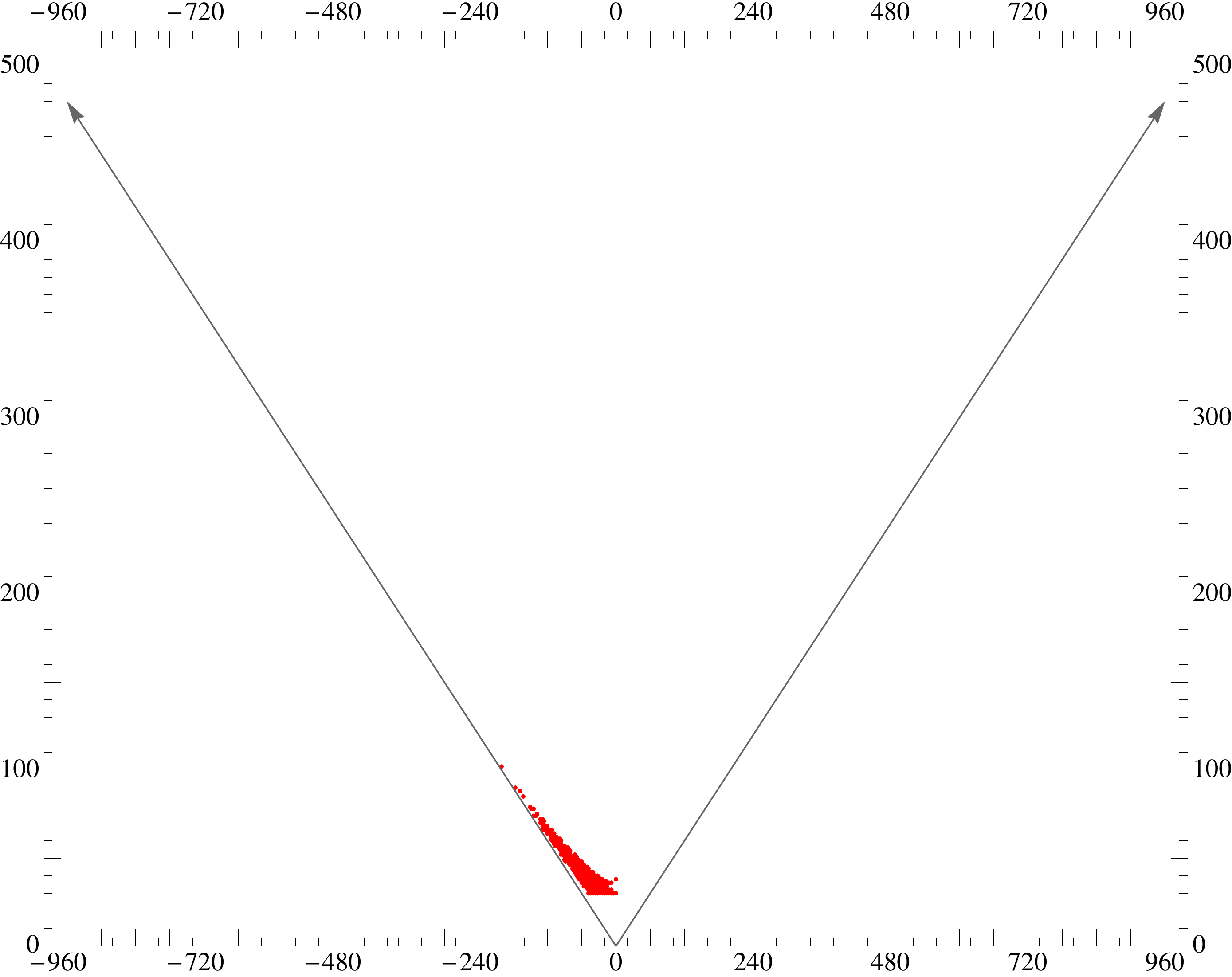}
\vskip10pt
\capt{5.7in}{CICYplot}{The Hodge numbers of the CICY manifolds. The horizontal axis corresponds to
$\chi=2(h^{11}-h^{21})$, while the vertical axis corresponds to $h^{11}+h^{21}$. While inappropriate to the CICY's the scale of the plot corresponds to the plots that come later.}
\end{center}
\end{figure}

There was a lot of work also, at about this time, on orbifolds. The algebraic constructions, such as the CICY's, tend to have many complex structure parameters (these correspond, roughly speaking, to the coefficients in the polynomials) and few \K parameters (again roughly, these correspond to the number of projective embedding spaces of a CICY). Thus for these spaces the Euler number, $\ch=2(h^{1,1}-h^{2,1})$, tends to be negative, and is in fact negative with $0\geq \ch\geq -200$ for all the CICY's. The orbifolds, by contrast, start with a torus, that has $\ch=0$, then a quotient is taken and the resulting singularities are resolved. The resolution introduces \K parameters and so increases the Euler number. Thus the orbifolds have positive Euler number. There was, for a while, a misguided debate as to whether there are more \cys with $\chi$ positive or $\chi$ negative.

A subsequent important construction involves weighted projective spaces. I learnt of these from a paper of Strominger and Witten~\cite{Strominger:1985it} and how to construct large classes from these spaces from a discussion with Brian Greene. In this construction one considers a weighted $\IP^4$ with coordinates $x=(x_1,\ldots,x_5)$ and weights $k=(k_1,\ldots,k_5)$ such that the coordinates are identified under a scaling
$$
\IP^4_k~:~~~(x_1,x_2,\ldots,x_5)\simeq (\l^{k_1} x_1, \l^{k_2}x_2,\ldots,\l^{k_5}x_5)~.
$$
The first thing to note is that a weighted projective space is an orbifold of an ordinary projective space. This is seen most easily by setting $x_j=y_j^{k_j}$, so that the scaling relation becomes
$(y_1,\ldots,y_5)\simeq \l (y_1,\ldots,y_5)$, however we must also identify $y_j\simeq \ee^{2\p\ii/k_j} y_j$.
Thus we~have
$$
\IP^4_k~=~\frac{\IP^4}{\IZ_{k_1}\!\!\times\ldots\times \IZ_{k_5}}~,
$$
so hypersurfaces in weighted projective spaces are a type of hybrid between the hypersurfaces in ordinary projective spaces and the orbifolds.
We may seek to construct CY hypersurfaces by choosing a polynomial, $p(x)$, homogeneous of degree $d=\sum_j k_j$, under the scaling. It seems that there will be many possible choices of weights, since we can choose the weights, and these determine $d$. There are indeed many ways to choose the weights, but the number is limited by the fact that, if strange weights are chosen, it can become difficult to form monomials of the required degree, and if there are not sufficiently many of these then the resulting polynomial will not lead to a smooth manifold. If one understands the polynomial as corresponding to the Landau-Ginzburg potential of a conformal field theory then it seems natural to impose the condition that the partial derivatives 
$\partial p/\partial x_j$ could only all vanish simultaneously if all the coordinates vanish. A simple choice of bad weights is $k=(1,1,1,1,5)$. With this choice, we must have a polynomial of degree 9. If we denote the four coordinates of weight 1 by $x_j$, $j=1,..,4$ and the coordinate of weight 5 by $y$ then the polynomial can only be of the form
$$
p(x,y)~=~R_9(x) + R_4(x)\, y~,
$$
with $R_9(x)$ a polynomial of degree 9 and $R_4(x)$ a polynomial of degree 4. It is clear that, for such a polynomial, all the derivatives of $p$ vanish at the point $(0,0,0,0,1)$. This example is an example of a set of weights that are genuinely bad. It turns out, however, that the criterion just given, that the derivatives of the polynomial can only vanish simultaneously at the origin, is too strong. In some cases the derivatives can be allowed to vanish simutaneously at other points also. The true criterion was not known until later.

Certain types of polynomials could be shown to satisfy the `Landau-Ginzburg criterion'. For example the polynomials of Fermat type 
\begin{align*}
p~&=~x_1^{n_1} + x_2^{n_2} + x_3^{n_3} + x_4^{n_4} + x_5^{n_5}~:~
&\includegraphics[width=1.5in]{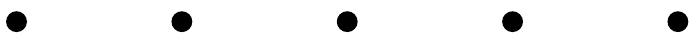}\\
\intertext{to which we have appended a graphical depiction.
Note, however, that a polynomial of Fermat type is only possible if $k_j$ divides $d$ for each $j$. There are about 800 sets of weights of this type. 
Other types of polynomial are also good, for example}
p~&=~x_1^{n_1} x_2 + x_2^{n_2} + x_3^{n_3} + x_4^{n_4} + x_5^{n_5}~:~
&\includegraphics[width=1.5in]{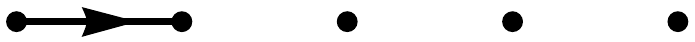}\\
\intertext{or}
p~&=~x_1^{n_1} x_2 + x_2^{n_2}x_3 + x_3^{n_3} + x_4^{n_4}x_5 + x_5^{n_5}x_4~:~
&\includegraphics[width=1.5in]{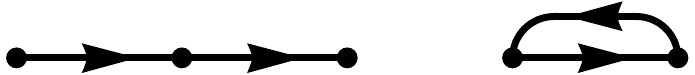}
\end{align*}

It is possible, as in the last polynomial above, to mix and match types. In this way a list of some 6,000 transverse polynomials was put together and the corresponding Hodge numbers 
calculated~\cite{Candelas:1989hd}. This list showed a very pleasing symmetry under the interchange $h^{1,1}\!\leftrightarrow\, h^{2,1}$. The symmetry was not perfect, since about 10\% of the pairs $\hodgenos$ did not have a mirror partner. On the other hand the classification of the polynomials was known to be incomplete. I should say that producing the list seemed at the time like a major enterprise that involved shipping parts of the computation to (what was at the time) a supercomputer. The problem was that for each type of polynomial it was necessary to find all the allowed polynomials, that is all the allowed powers $(n_1,\ldots,n_5)$. For each type of polynomial there are upper bounds on the powers $n_j$ but the volume of the space of possible powers is very large and to search this in reasonable time requires sophisticated programming.
\begin{figure}[!t]
\begin{center}
\includegraphics[width=6.5in]{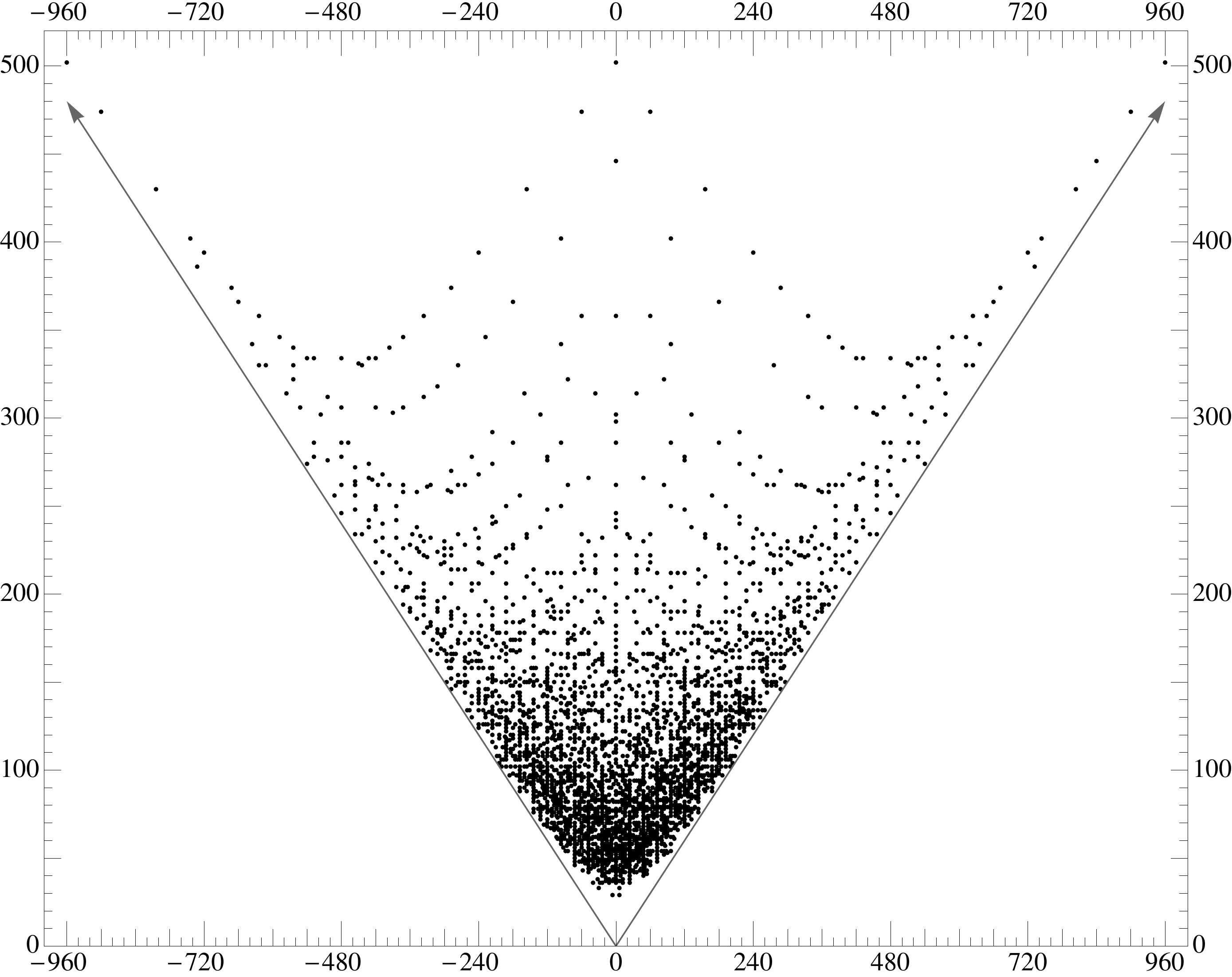}
\capt{5.5in}{WP4plot}{The Hodge numbers of the list of 7,555 transverse polynomials in weighted $\IP^4$'s, that was calculated by Kreuzer and Skarke and also by Klemm and Schimmrigk. The original plot of \cite{Candelas:1989hd} was somewhat less dense and somewhat more symmetric under the interchange $h^{11}\leftrightarrow h^{21}$. Again, The horizontal axis corresponds to $\chi=2(h^{11}-h^{21})$, while the vertical axis corresponds to $h^{11}+h^{21}$.}
\end{center}
\end{figure}

I have gone into some detail because this is when I first came to know Max. Rolf Schimmrigk and Albrecht Klemm~\cite{Klemm:1992bx} and, independently, Max and Harald Skarke~\cite{Kreuzer:1992da} had taken on the task of computing the complete list of types of transverse polynomials. Both collaborations worked away and had to understand and implement a procedure for listing all types of allowed polynomials, of which there are about 1,000 types, and for each find all polynomials of the given type. Both collaborations announced a solution within 24 hours of each other. Klemm and Schimmrigk announced a list of $7,555$ polynomials and Kreuzer and Skarke announced a list of $7,554$ transverse polynomials. This was the only fact that was initially communicated, not the list of the polynomials themselves. Such was the sophistication of their understanding that, given this single fact, Max and Harald immediately found the polynomial, of a very degenerate type, that they were missing, and with this addition the two lists were identical. This I think illustrates what characterised Max's work: a combination of physical insight, sophisticated mathematics and a rare skill with programming. 

These 7,555 polynomials give 2997 different pairs of Hodge numbers. These give rise to \fref{WP4plot} which is similar to the original figure of  \cite{Candelas:1989hd} though somewhat denser. Surprisingly the left-right asymmetry of the plot increased as the result of doing a more complete analysis. 

Victor Batyrev~\cite{Batyrev:1993dm} had recently made his seminal observations about the relationship between reflexive polyhedra and \cys that can be realised as hypersurfaces in toric varieties. Victor had come to CERN to give a seminar which, owing to the fact that toric geometry was a completely mysterious subject to physicists at the time, was not, it is fair to say, fully understood by the audience.

Xenia and I were intrigued by the polynomials that did not have a mirror and were corresponding with Sheldon Katz about this. Sheldon did understand toric geometry and Batyrev's construction and was patiently explaining how the polynomials in weighted projective space might give rise to reflexive polyhedra and, if so, have a mirror in the sense of Batyrev. Checking this required writing code. Typically a weighted polynomial, of the sort we are discussing, will admit, say, two hundred monomials of the requisite degree. Such a monomial is of the form $x_1^{m_1} x_2^{m_2}\ldots x_5^{m_5}$ and corresponds to a vector $m=(m_1,m_2,\ldots,m_5)$. The $m$'s have five components but all have the property that 
$m\cdot k=d$, where $k$ is the weight vector. So the monomials live, in fact, in a four dimensional vector space and we have to check whether they form a reflexive polyhedron. The monomials certainly form a four dimensional polyhedron and the first task is to locate the vertices. This is a task which is, in general, impractical without a computer. In any event together with Sheldon we were able to check that all the $7,555$ weighted polynomials was associated to a reflexive polyhedron in the sense of Batyrev. 

Max and Harald, seeing that Batyrev's procedure provides a way of extending the interpretation of the polynomials that is more general than understanding them as Landau-Ginzburg potentials, then embarked on a project of many parts \cite{Kreuzer:1995cd,Kreuzer:2000qv, Kreuzer:2000xy, Kreuzer:2002uu}, whose successful conclusion I regard as a tour de force of mathematical physics and computation, which was to list all four dimensional reflexive polyhedra, of which there turned out to be almost 500,000,000. The number of reflexive polyhedra is very large. The difficulty is not so much this, as that in order to find polyhedra that are reflexive you have to construct many more that are not reflexive and search for the reflexive ones among these. Even for the problem of listing the weighted polynomials, the theoretical part of listing all types of allowed polynomials was only part of the problem, as noted previously. 
\begin{figure}[!t]
\begin{center}
\includegraphics[width=6.5in]{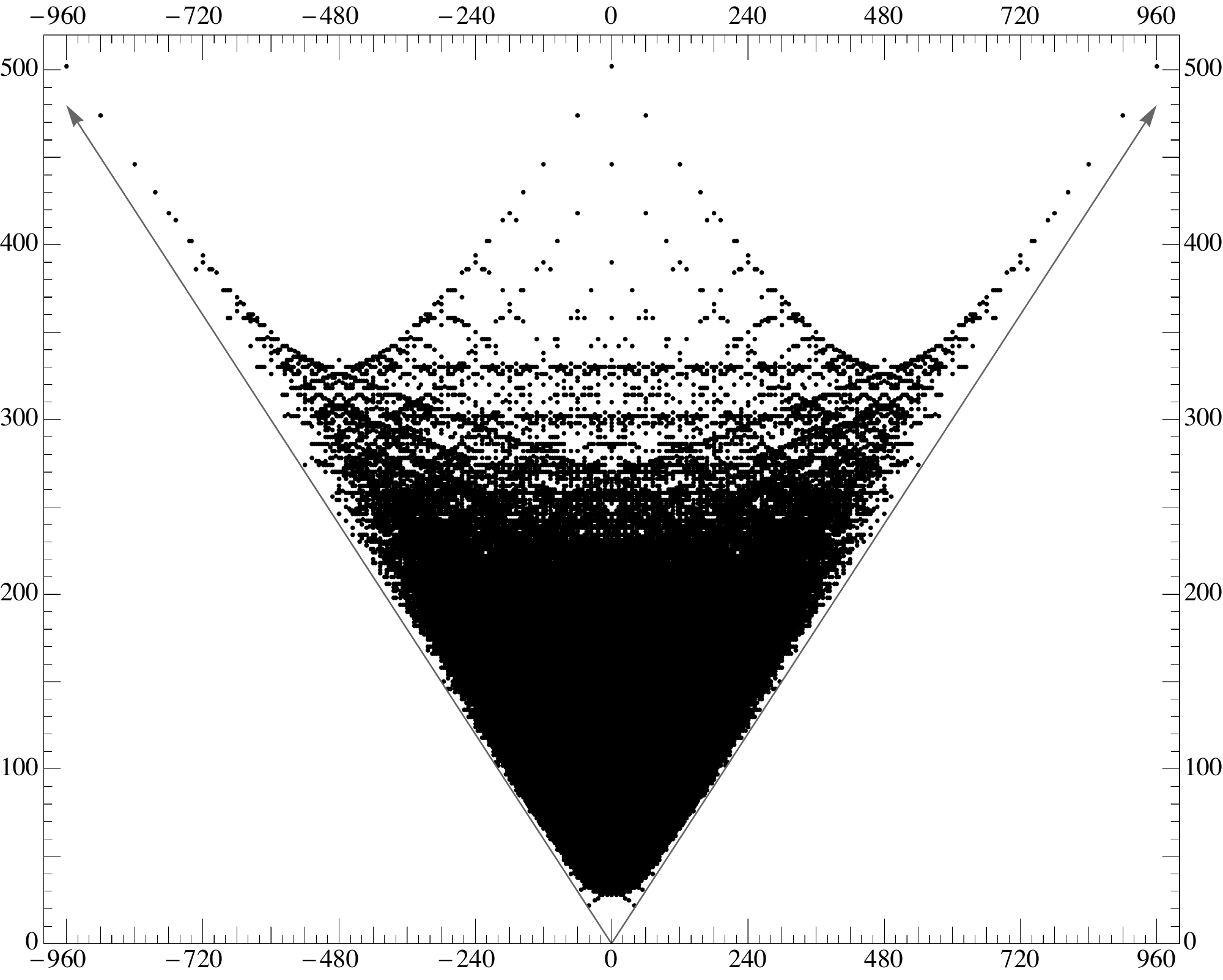}
\vskip0pt
\capt{6.0in}{BasicKSplot}{The  30,108 pairs of Hodge numbers for the polyhedra of the Kreuzer-Skarke list. }
\end{center}
\vspace*{-15pt}
\end{figure}

Given a reflexive polyhedron one needs to analyse it by finding the faces of each dimension and counting the number of monomials that lie in each. From this information one calculates the Hodge numbers. This is straightforward, in principle, but remember that this must be done $5{\times}10^8$ times and that a year has only $3{\times}10^7$ seconds. Max and Harald overcame these problems and the result is the list behind the Kreuzer-Skarke plot of~\fref{BasicKSplot}. How do we know that Max and Harald constructed the list correctly? Max and Harald constructed each reflexive polyhedron, each such polyhedron has a reflexive dual, but this fact was not used in the construction. So if one polyhedron had been missed in the construction the list would not have been symmetric under the operation of replacing each polyhedron by its dual, which it is. To show the symmetry it would have to be the case that at least two polyhedra had been missed and that these are duals of each other. The probability of this having happened, given that the construction is independent of the duality is astronomically small. 

What will be done with this list? Two projects, one that has been carried through for the CICY's, and one that is in the process of being implemented for the CICY's, have important extensions to the Kreuzer-Skarke list, if only one can understand the geometry sufficiently well and solve the problem of calculating efficiently. 

Recently, Volker Braun completed a computer scan of the CICY's and found all linear, fixed point free automorphisms. The quotients by these freely acting symmetries give rise to a number of remarkable \cys with small Hodge numbers that live in the tip of the distribution; so in the region of the plot where both Hodge numbers are small. These are in some cases remarkably symmetric spaces that are also, in some sense, the simplest \cys. There is some overlap between the CICY's and the Kreuzer-Skarke list and an extension of the techniques used for the CICY's can be implemented in the toric case. As an example of what can be done in this way, Braun has constructed a \cym with Hodge numbers $\hodgenos=(1,1)$ as the quotient of the manifold corresponding to the 24-cell, which is one of the polyhedra of the Kreuzer-Skarke list. This construction is reported in the article by Braun in this volume. 

The second project, is one that has been in development for a number of years that involves Andr\'e Lukas and his collaborators, is searching for holomorphic vector bundles on the CICY's that give rise to phenomenologically viable heterotic vacua (see, for example, \cite{anderson:2012yf} and references cited therein). The history of this project has again been one of overcoming the combinatoric obstacles to performing searches in reasonable time. One would hope to extend these searches to the hypersurfaces in toric varieties, or at least, in the first instance, to those with small Hodge numbers. The extension of this work to hypersurfaces in toric varieties has begun~\cite{he:2009wi, he:2011rs}, and it fitting that Max should be a co-author on this last paper.

Max's work on \cys was always motivated by a desire to understand the vacua of string theory. Thus Max had an enduring interest in heterotic string vacua and the closely allied subject of F-theory. On heterotic string theory there was a series of four papers on the `UPenn'  
model~\cite{Braun:2007tp, Braun:2007xh, Braun:2007vy, Braun:2008sf}, that are described in the contribution of Ovrut to this volume. Quite apart from particular models, Max was keenly interested in the bridge between the conformal field theories, corresponding to string theory vacua, and the language of geometry, manifolds and vector bundles and wrote on this in~\cite{Kreuzer:2009ha, Kreuzer:2010ph}, for example. On the subject of 
F-theory there were papers with Knapp and co-workers~\cite{Knapp:2011ip, Knapp:2011wk, Chen:2010ts}.
Work on extending the constructions of \cys beyond hypersurfaces in toric varieties included a consideration of complete intersections in toric varieties~\cite{Kreuzer:2001fu, Klemm:2004km} and work with Batyrev~\cite{Batyrev:2008rp, Kreuzer:2009is} on a construction of \cy threefolds via conifold transition from varieties described by reflexive polyhedra. 

The `compactification' of string theory on a manifold, or in the case of heterotic string theory, a vector bundle together with a manifold, has proved to be a surprisingly durable process that has formed an essential part of our understanding of string theory. This has been so since the first widespread interest in string theory. As long as this remains an essential process, it is equally essential to be able to describe the manifold on which the theory is compactified. It is this fact that Max understood so well and to which he devoted so much of his research.
\newpage
\section*{Acknowledgements}
I am grateful to Harald Skarke and Andrei Constantin for their comments on a draft of this article and to The Abdus Salam International Center for Theoretical Physics for hospitality and support.

\vspace*{20pt}
\bibliographystyle{utphys}
\bibliography{BibliographyPC}
\end{document}